\documentclass[aip,reprint]{revtex4-2}

\usepackage[utf8]{inputenc}
\usepackage[T1]{fontenc}
\usepackage{mathptmx}       % Times font for math + text
\usepackage{graphicx}       % Figures
\usepackage{dcolumn}        % Decimal-aligned columns
\usepackage{bm}             % Bold math symbols
\usepackage{makecell}       % Table enhancements
\usepackage{xcolor}         % Colors
\usepackage{needspace}
\usepackage{amsmath}
\usepackage{amssymb}
\usepackage{float}
\usepackage{placeins}
\usepackage{subcaption}     % For subfigures (better than subfig)

\usepackage{hyperref}
\hypersetup{colorlinks=true, linkcolor=blue, citecolor=blue, urlcolor=blue}

\usepackage[normalem]{ulem} % Underline / strikeout
\renewcommand{\emph}[1]{\textit{#1}} % Ensure emph = italics

\begin{document}

\preprint{AIP/123-QED}

\title[]{\textcolor{black}{Eulerian–Lagrangian relations in decaying two-dimensional incompressible Navier-Stokes fluids across initial vorticity packing and Reynolds number}}

% Force line breaks with \\
\author{ Snehanshu Maiti}
\email{snehanshu.maiti@gmail.com}

\affiliation{%
Institute for Plasma Research, Bhat, Gandhinagar, Gujarat 382428, India%\\This line break forced% with \\
}%

%\author{ Shishir Biswas}%
% \email{Second.Author@institution.edu.}
%\affiliation{ 
%Department of Physics and Astronomy, University of Notre Dame, Notre Dame, IN \textcolor{black}{46556}, USA%\\This line break forced with \textbackslash\textbackslash
%}%

\author{Rajaraman Ganesh}
% \homepage{http://www.Second.institution.edu/~Charlie.Author.}
\affiliation{%
Institute for Plasma Research, Bhat, Gandhinagar, Gujarat 382428, India%\\This line break forced% with \\
}%
\affiliation{%
Homi Bhabha National Institute, Training School Complex, Anushaktinagar, Mumbai 400094, India.%\\This line break forced% with \\
}%

\date{\today}% It is always \today, today,

\begin{abstract}
 
\textcolor{black}{{Recent studies [“Vorticity packing effects on long-time turbulent transport in decaying two-dimensional incompressible Navier–Stokes fluids,” Phys. Fluids 38, 045159 (2026)] demonstrated that, at a fixed high Reynolds number (Re), the initial vorticity packing fraction (VPF) governs the coupled Eulerian flow evolution and Lagrangian tracer particle transport in decaying two-dimensional incompressible Navier–Stokes fluids, revealing a strong Eulerian–Lagrangian relationship during the nonequilibrium inverse-cascade regime and a direct Eulerian–Lagrangian correspondence in the late-time coherent-vortex quasi-equilibrium regime,  wherein increasing VPF drives transitions from point-vortex to patch-vortex equilibria and from subdiffusive to superdiffusive transport. In the present work, we investigate how these Eulerian–Lagrangian connections evolve across a broad (VPF, Re) parameter space. The results show that the Eulerian–Lagrangian relationship remains largely preserved during the nonequilibrium inverse-cascade regime, where transport increases systematically with VPF and remains primarily controlled by VPF despite secondary Re-dependent oscillatory modulation. In contrast, the late-time coherent-vortex quasi-equilibrium regime exhibits Eulerian statistical equilibria that remain largely insensitive to Re, while the corresponding tracer-particle transport displays a strong Re dependence characterized by strong oscillatory and non-monotonic variations across the (VPF, Re) parameter space, substantially weakening the VPF-ordered transport hierarchy observed in the inverse-cascade regime. Consequently, for the parameter range, spatial resolutions, and integration times explored in the present study, the direct Eulerian–Lagrangian correspondence identified at fixed high Re is not universally maintained across the broader (VPF, Re) parameter space.}}

\end{abstract}

\maketitle

\section{Introduction}

A recent study by Maiti \textit{et al.}, Phys. Fluids \textbf{38}, 045159 (2026) (referred to henceforth as MBG2026), employed high-resolution direct numerical simulations (DNS) to investigate the influence of the \textcolor{black}{initial} vorticity packing fraction (VPF), which quantifies the initial fractional area occupation of nonzero vorticity, on the coupled evolution of Eulerian flow organization and associated Lagrangian transport statistics in decaying two-dimensional incompressible Navier--Stokes fluids \textcolor{black}{at a fixed high Reynolds number\cite{Maiti2026}}. The study demonstrated a strong Eulerian--Lagrangian relationship during  the nonequilibrium inverse-cascade regime and a direct Eulerian--Lagrangian correspondence in the late-time quasi-equilibrium coherent-vortex regime, with VPF acting as the primary control parameter governing the transition from loosely packed low-VPF states to tightly packed high-VPF states. During the inverse-cascade regime, increasing VPF \textcolor{black}{was found to enhance} vortex interaction, merger activity, the persistence time of the inverse cascade, and the transition from anisotropic to more isotropic turbulence, while the associated Lagrangian transport progressively transitioned from subdiffusive to superdiffusive behavior \textcolor{black}{with increasing VPF}. At late times, the Eulerian statistical states transitioned from the point-vortex sinh--Poisson regime to the finite-sized patch-vortex \textcolor{black}{Kuzmin--Miller--Robert--Sommeria (KMRS) regime\cite{BG2022}} with increasing VPF, while the corresponding Lagrangian transport again exhibited a transition from subdiffusive to superdiffusive dynamics\cite{Maiti2026}. Together, these results established clear Eulerian–Lagrangian connections between flow organization and the underlying transport dynamics across different stages of turbulent evolution. 
\textcolor{black}{The present study extends MBG2026 by examining how these Eulerian–Lagrangian connections evolve across a broader two-parameter space spanning different VPF and Re.
}

We use the coupled GPU-accelerated two-dimensional incompressible hydrodynamic and tracer-particle solver, \textcolor{black}{GHD2D--Particle \cite{Maiti2026,BG2022},} developed in-house at the Institute for Plasma Research, to solve the governing fluid equations in a doubly periodic square domain while simultaneously evolving passive tracer particles in the resulting flow field. The numerical framework and simulation setup have been extensively benchmarked and validated for both fluid and tracer-particle dynamics. Two-dimensional incompressible Navier--Stokes turbulent flows are generated from Kelvin--Helmholtz unstable shear layers, which subsequently undergo inverse cascade and self-organize into large-scale coherent vortices at late-time quasi-equilibrium states. The turbulence properties are systematically varied through changes in VPF and Re. The associated Lagrangian absolute dispersion is characterized using the mean-square displacement (MSD) and the corresponding transport exponents ($\alpha$) of tracer particles across the inverse-cascade and long-time coherent-vortex regimes; \textcolor{black}{both quantities are defined in Sec. II.} Thus, we investigate how the Eulerian--Lagrangian connections across different stages of turbulence evolve over a broad range of VPF and Re.

In Section II, we describe the governing equations, numerical solvers, simulation setup, and diagnostic methods employed in this study. Section III presents our results on the evolution of nonequilibrium inverse-cascade dynamics, late-time quasi-equilibrium coherent-vortex states, and the associated Lagrangian transport behavior together with their dependence on VPF and Re. Finally, in Section IV, we summarize the main findings of the study and discuss how the Eulerian--Lagrangian relationship in the nonequilibrium inverse-cascade regime and the Eulerian--Lagrangian correspondence in the late-time quasi-equilibrium regime evolve across the broader VPF and Re parameter space.

\needspace{2\baselineskip}

\section{Simulations and Diagnostics}

\noindent\textbf{Eulerian Navier--Stokes fluid and Lagrangian tracer particle dynamics}\\

We describe here the fluid and particle equations, numerical solvers, and simulation setup employed to generate decaying two-dimensional turbulence fields over a range of \textcolor{black}{initial} vorticity packing fractions (VPF) and Reynolds numbers (Re), while simultaneously tracking tracer-particle trajectories to quantify transport in each case.

The dynamics of decaying two-dimensional incompressible Navier--Stokes flow is governed by the dimensionless vorticity ($\omega$)--streamfunction ($\psi$) formulation,

\begin{equation}
\frac{\partial \omega}{\partial t}
=
[\psi,\omega]
+
\frac{1}{\textcolor{black}{\mathrm{Re}}}\nabla^2\omega,
\qquad
\omega = -\nabla^2\psi.
\label{eq:navier_stokes_scalar}
\end{equation}

Here, $[\psi,\omega]
=
\partial_x\psi\,\partial_y\omega
-
\partial_y\psi\,\partial_x\omega$
denotes the Poisson bracket, which represents the nonlinear advection of vorticity by the flow. The Reynolds number, $\mathrm{Re}$, characterizes the ratio of inertial to viscous forces, with larger values corresponding to increasingly turbulent flows and weaker viscous dissipation. The Lagrangian description involves tracking passive tracer particles that move with the flow but do not influence the fluid dynamics. The motion of such particles is governed by the following equations:

    \begin{equation}
        \frac{dx}{dt}= \frac{\partial{\psi}}{\partial{y}}; \quad \quad 
        \frac{dy}{dt}=   - \frac{\partial{\psi}}{\partial{x}}
         \label{eq:adv_eqn}
        \end{equation} 

\hspace{-0.33cm}Here, \( x(t) \) and \( y(t) \) represent the instantaneous positions of the tracer particle as functions of time, whose evolution is determined from the streamfunction.

The coupled GPU-accelerated hydrodynamic and tracer-particle framework, GHD2D--Particle, is employed to numerically evolve the two-dimensional incompressible Navier--Stokes equations in a doubly periodic square domain together with passive tracer particle trajectories in the resulting flow field. The fluid solver employs a GPU-accelerated pseudospectral Fourier method, with nonlinear terms evaluated using FFTs, standard 2/3 dealiasing \cite{Patterson1971}, and second-order Adams--Bashforth time integration \cite{Adams1855, Bashforth1883}, implemented using the CUDA-based cuFFT library \cite{nvidia_cufft}. Particle trajectories are integrated using a fourth-order Runge--Kutta (RK4) scheme \cite{Press2007}, with particle positions obtained through linear interpolation of the streamfunction field. Previous studies by MBG2026 demonstrated that trajectories obtained using linear interpolation are quantitatively similar to those obtained using higher-order cubic interpolation of both the velocity and streamfunction fields \textcolor{black}{due to the high grid resolution employed in the simulations}; therefore, the present study employs linear interpolation of the streamfunction throughout. The numerical framework and simulation setup have been benchmarked against standard Kelvin–Helmholtz instability growth rates for the fluid solver and passive tracer transport dynamics in a two-dimensional oscillating Taylor--Green flow for the particle solver \cite{Maiti2026}.

The flow is initialized using alternating-sign vorticity strips to generate Kelvin--Helmholtz instabilities, which subsequently evolve into freely decaying two-dimensional incompressible Navier--Stokes turbulence, exhibiting inverse-cascade dynamics and long-time  quasi-equilibrium coherent vortices. We generate various regimes of freely decaying two-dimensional incompressible turbulence using the GHD2D--Particle framework at a grid resolution of $2048^2$ by systematically varying the  initial vorticity packing fraction, controlled through the number of alternating-sign vorticity strips in the domain, and the Reynolds number, thereby constructing a two-parameter (VPF, \textcolor{black}{Re}) space of turbulent dynamics and transport. The computational domain is a doubly periodic square of size $L=2\pi$, with each strip having width $d=\pi/16$, while the regions between strips remain non-circulating. The corresponding \textcolor{black}{initial} VPF is defined as
\begin{equation}
VPF = \frac{ndL}{L^2},
\end{equation}

\begin{figure*}
    \centering
    \begin{subfigure}[b]{0.245\textwidth}
        \centering
         \includegraphics[width=5.5cm]{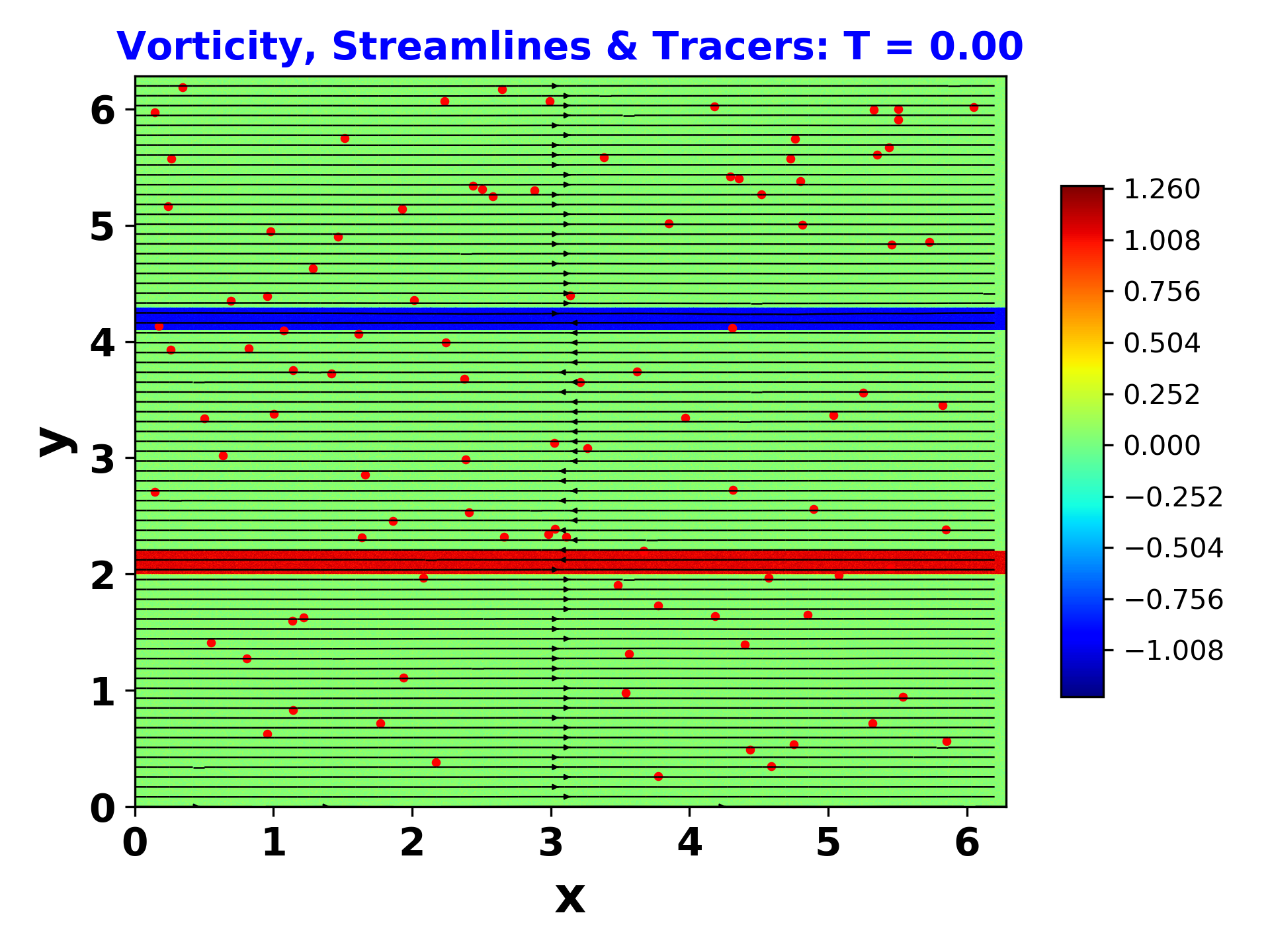}
        \caption{\textbf{2 strips}}
 %       \label{aa}
    \end{subfigure}
    \centering
    \begin{subfigure}[b]{0.245\textwidth}
        \centering
         \includegraphics[width=5.5cm]{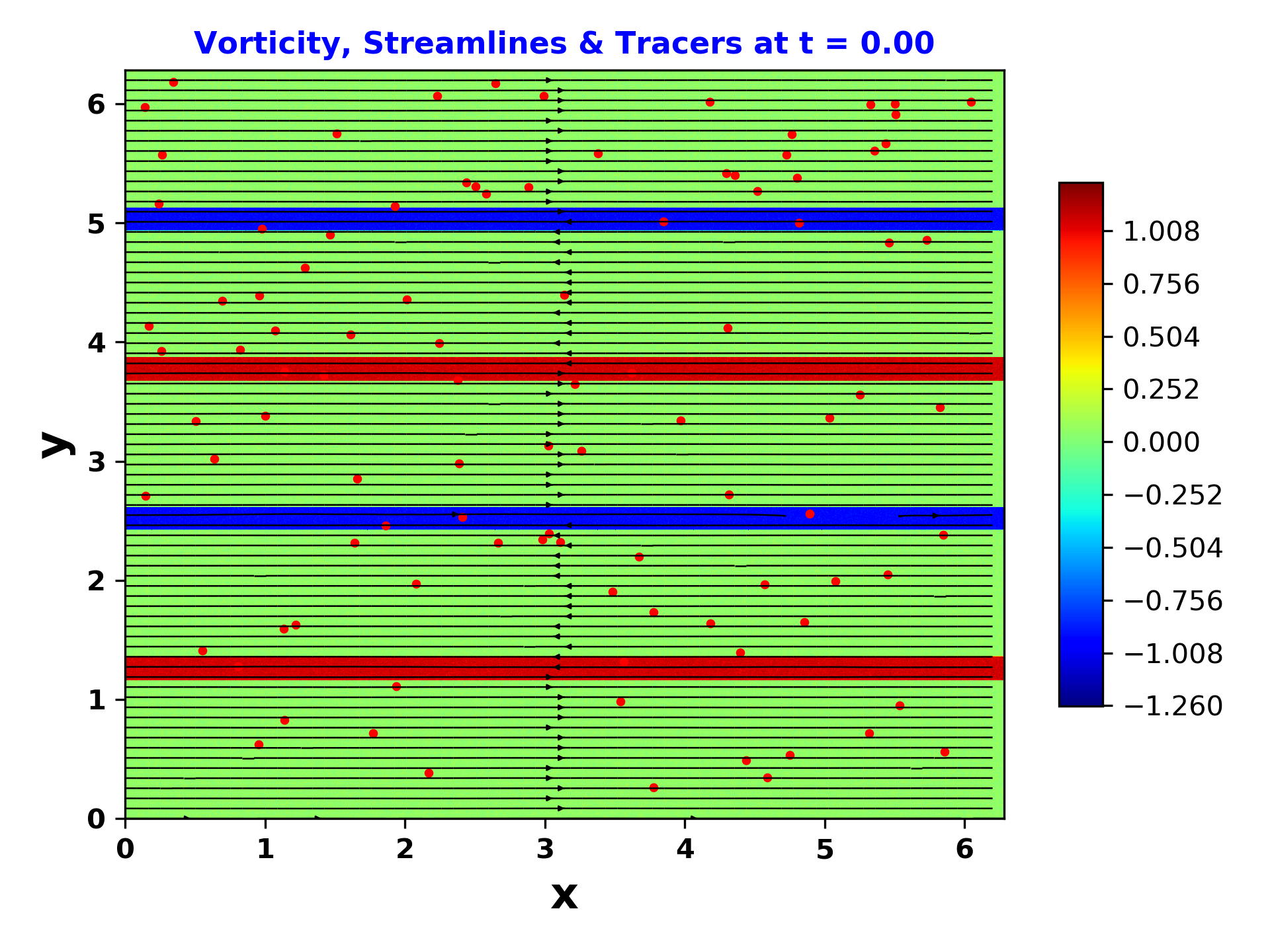}
        \caption{\textbf{4 strips}}
  %      \label{aa}
    \end{subfigure}
    \centering
    \begin{subfigure}[b]{0.245\textwidth}
        \centering
         \includegraphics[width=5.5cm]{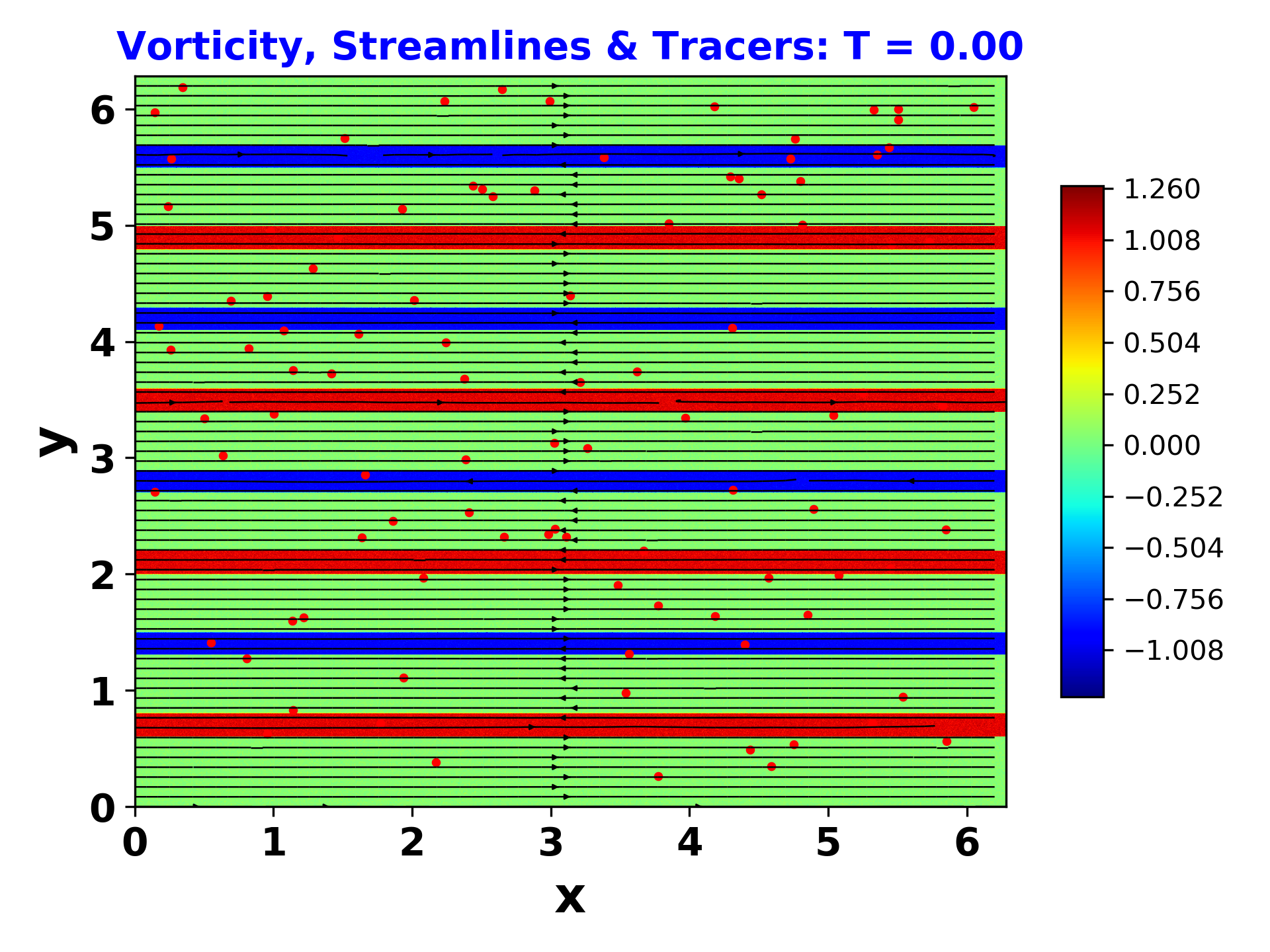}
        \caption{\textbf{8 strips}}
 %       \label{aa}
    \end{subfigure}
    \centering
    \begin{subfigure}[b]{0.245\textwidth}
        \centering
         \includegraphics[width=5.5cm]{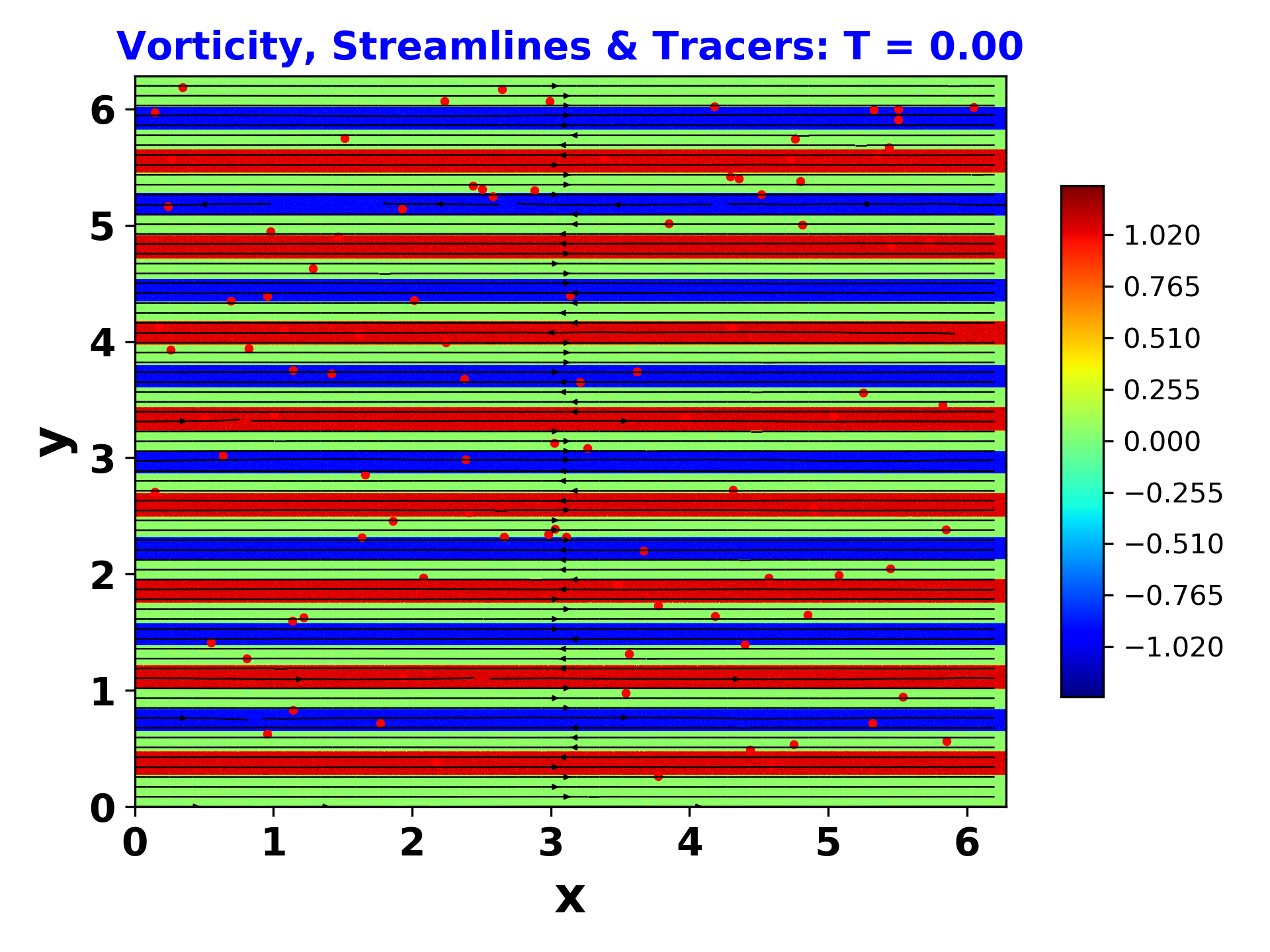}
        \caption{\textbf{16 strips}}
  %      \label{aa}
    \end{subfigure}
   \caption{ Initial vorticity, stream function and associated tracer particles distribution at \textcolor{red}{time}, $T$ = 0 for various cases of turbulence with strip-based initial conditions in our numerical simulations. (a) 2 strip configuration with lowest vorticity packing fraction (VPF 6.25\%) (b) 4 strip configuration with low VPF (12.5\%)   (c) 8 strip configuration with moderate VPF (25\%)  (d) 16 strip configuration with high VPF (50\%). The vorticity value for blue, red and green regions are -1, 1 and 0 respectively.}
\label{fig:IC}
\end{figure*}

\hspace{-0.4cm}\textcolor{black}{where} $n$ is the number of strips, corresponding to VPF values of $6.25\%$, $12.5\%$, $25\%$, $50\%$, and $62.5\%$ for $n=2,4,8,16$, and $20$, respectively (see Fig.~\ref{fig:IC}). The initial configuration is circulation neutral, with equal positive and negative circulations. For pseudospectral simulations employing standard $2/3$ dealiasing, the maximum dynamically resolvable wavenumber scales as $k_{\max}=N_{\mathrm{grid}}/3$ and the corresponding maximum numerically resolvable Reynolds number scales as Re$_{\max}\sim N_{\mathrm{grid}}^2$, yielding Re$_{\max} = 4194304$ for the present simulations performed at a spatial resolution of $2048^2$. Accordingly, the present study investigates a broad range of Reynolds numbers, Re = 10000, 114288, 210000, 228576, 240000, 457152, 914304, 1828608, 4194304, spanning weakly to strongly turbulent regimes while remaining within the numerical resolution limits of the simulations. To trigger Kelvin--Helmholtz instability and the subsequent turbulent evolution, a small-amplitude random perturbation is added to the vorticity field:
\begin{equation}
\delta \omega = \sum_{m=1}^{64} \alpha \cos(mx+\phi_m),
\end{equation}

\hspace{-0.4cm}\textcolor{black}{where} $\alpha=0.01$, $m$ is the mode number, and $\phi_m$ are random phases uniformly distributed between $-\pi$ and $\pi$.  The coupled fluid and tracer-particle dynamics are evolved using a fixed time step of $\Delta t = 10^{-3}$  up to  $T$ = 3000, ensuring the system reaches a long-time quasi-steady turbulent state for analysis. A total of 1000 tracer particles are used to characterize the transport and dispersion properties. Previous studies by MBG2026 performed particle-number convergence tests at full spatial resolution \textcolor{black}{of $2048^2$} and demonstrated that this choice is sufficient to obtain statistically converged and accurate Lagrangian transport measures.\\

\textbf{Fluid and particle transport diagnosis}\\

The late-time Eulerian statistical equilibria, derived from entropy–maximization principles, relate vorticity $(\omega)$ and stream function $(\psi)$ as follows.
For the point-vortex model (sinh–Poisson formulation)\cite{Montgomery1992}:
\begin{equation}
\omega_{\mathrm{PV}} = \alpha \sinh(-\beta \psi),
\label{eq:pv}
\end{equation}
and for the finite-size patch-vortex model (KMRS formulation)\cite{Kuzmin1982,Miller1990,Robert1991}:
\begin{equation}
\omega_{\mathrm{KMRS}} =
\frac{A e^{-B\psi} - C e^{B\psi}}{1 + [A e^{-B\psi} + C e^{B\psi}]}.
\label{eq:kmrs}
\end{equation}

It has been demonstrated that\cite{BG2022}, while at low vortex packing fractions (VPF) the late-time Eulerian statistical equilibria agree well with the point-vortex (sinh–Poisson) theory, at high VPF they conform more closely to the finite-size or patch-vortex (KMRS) theory. We investigate how these late-time statistical equilibria evolve with Reynolds number across different vortex packing fractions.

The transport is quantified through the time evolution of the ensemble-averaged mean-square displacement (MSD) of tracer-particle trajectories. The directional and total MSDs are defined as
\begin{equation}
\langle \Delta x^2(t) \rangle = \langle [x(t)-x_0]^2 \rangle, \quad
\langle \Delta y^2(t) \rangle = \langle [y(t)-y_0]^2 \rangle,
\end{equation}
\begin{equation}
\langle \Delta r^2(t) \rangle
=
\langle \Delta x^2(t) + \Delta y^2(t) \rangle,
\label{eq:msd}
\end{equation}

\hspace{-0.4cm}\textcolor{black}{where} \(x_0\) and \(y_0\) denote the initial particle positions, and \(\langle \cdot \rangle\) represents an ensemble average over all tracer particles. The MSD typically exhibits a power-law scaling,
\begin{equation}
\langle \Delta r^2(t) \rangle \propto t^{\alpha},
\end{equation}
\textcolor{black}{where} \(\alpha\) is the transport exponent characterizing the nature of particle transport. Values of \(\alpha<1\), \(\alpha=1\), and \(\alpha>1\) correspond to subdiffusive, diffusive, and superdiffusive transport, respectively.

\textcolor{black}{To distinguish transport behavior in the two dynamical regimes identified in the flow evolution, we introduce \(\alpha_T\) and \(\alpha_C\) as the MSD scaling exponents measured in the turbulent inverse-cascade regime and the late-time coherent-vortex regime, respectively. Furthermore, \(\langle \alpha_T \rangle_{\mathrm{Re}}\) and \(\langle \alpha_C \rangle_{\mathrm{Re}}\) denote the corresponding Reynolds-number-averaged transport exponents at fixed VPF, obtained by averaging \(\alpha_T\) and \(\alpha_C\), respectively, over all Reynolds numbers considered for that VPF.}

The corresponding directional and total diffusion coefficients are defined as
\begin{equation}
D_x = \textcolor{black}{\lim_{t\to\infty}}\frac{\langle \Delta x^2(t) \rangle}{2t},
\qquad
D_y = \textcolor{black}{\lim_{t\to\infty}}\frac{\langle \Delta y^2(t) \rangle}{2t},
\qquad
D = D_x + D_y.
\label{eq:diff_coeff}
\end{equation}

We investigate how the transport exponents,  \(\alpha_T\)  and \(\alpha_C\), vary with vortex packing fraction and Reynolds number in the nonequilibrium inverse-cascade turbulence regime and the late-time quasi-equilibrium coherent-vortex regime, respectively.

\section{Results}

In this section, we present results from numerical experiments exploring long-time turbulence evolution and tracer transport across the two-parameter (VPF, \textcolor{black}{Re})  space. We focus on  the inverse-cascade turbulence regime, and the long-time quasi-steady coherent dipole-motion phase. We examine the Eulerian equilibrium states using statistical-mechanical descriptions through $\omega-\psi$ correlations, and characterize transport through the time evolution of the mean-square displacement (MSD) and diffusion coefficients of tracer particles. We further investigate how Eulerian–Lagrangian connections evolve across the (VPF, \textcolor{black}{Re}) parameter space.
 
\subsection {Eulerian Dynamics}

The Eulerian fluid dynamics of turbulence and the associated late-time quasi-equilibrium statistical mechanics for various VPF values at a Reynolds number of Re = 228576 were presented in MBG2026. Here, we extend the analysis over a two-dimensional VPF and Re parameter space.

\begin{figure}[htbp]
    \centering
    \includegraphics[width=0.45\textwidth]{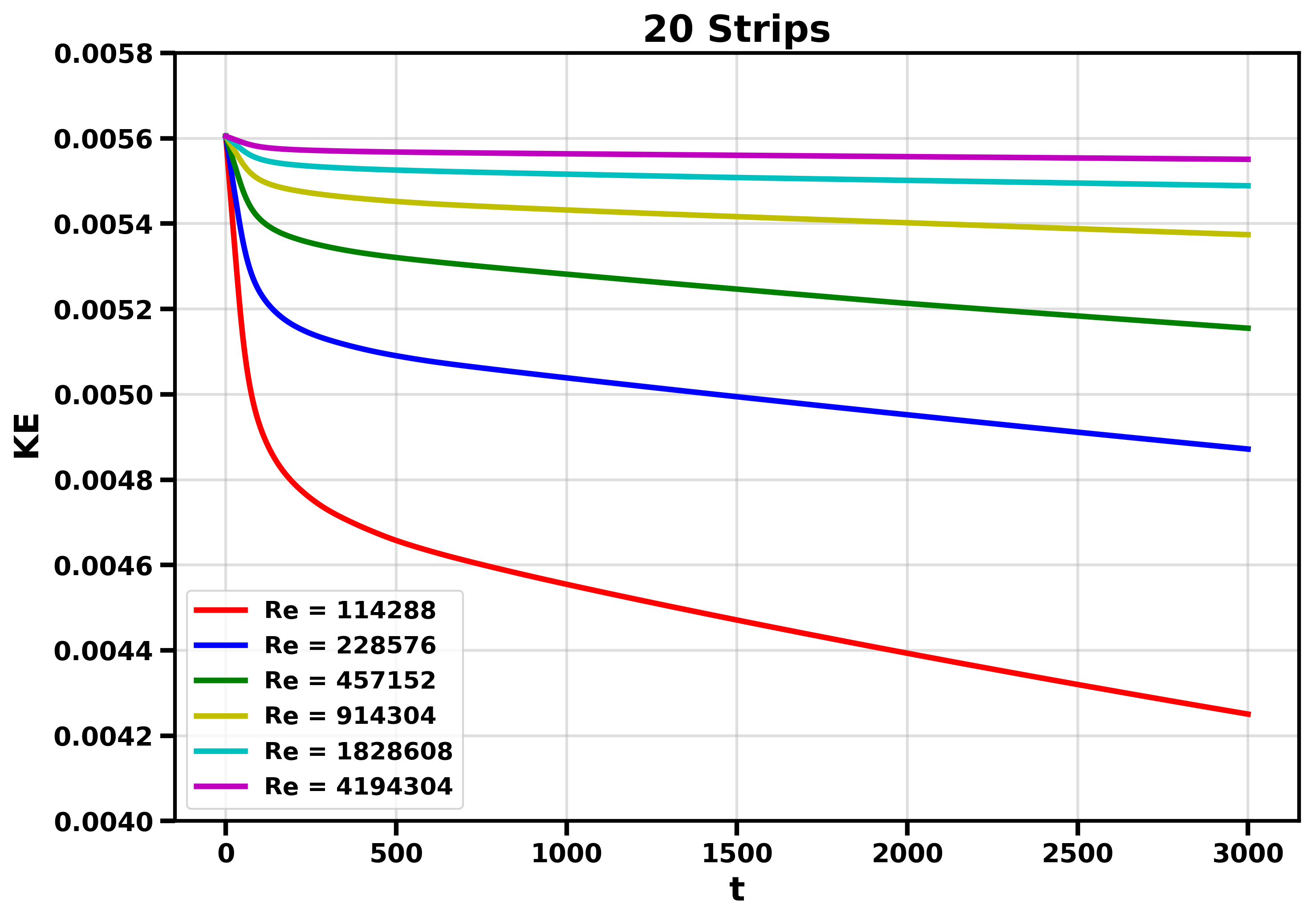}
    \caption{Kinetic-energy \textcolor{black}{(KE)} decay for the highest VPF case ($62.5\%$) across different Reynolds numbers.}
    \label{fig:KE}
\end{figure}

We first present \textcolor{black}{the kinetic-energy (KE) decay of turbulence} for the highest VPF case across different Re in Fig.~\ref{fig:KE}. As expected, lower-Re flows exhibit a more rapid decay due to enhanced viscous dissipation, leading to substantially different energy-evolution histories.

\begin{figure*}[htbp]
    \centering
    \includegraphics[width=\textwidth]{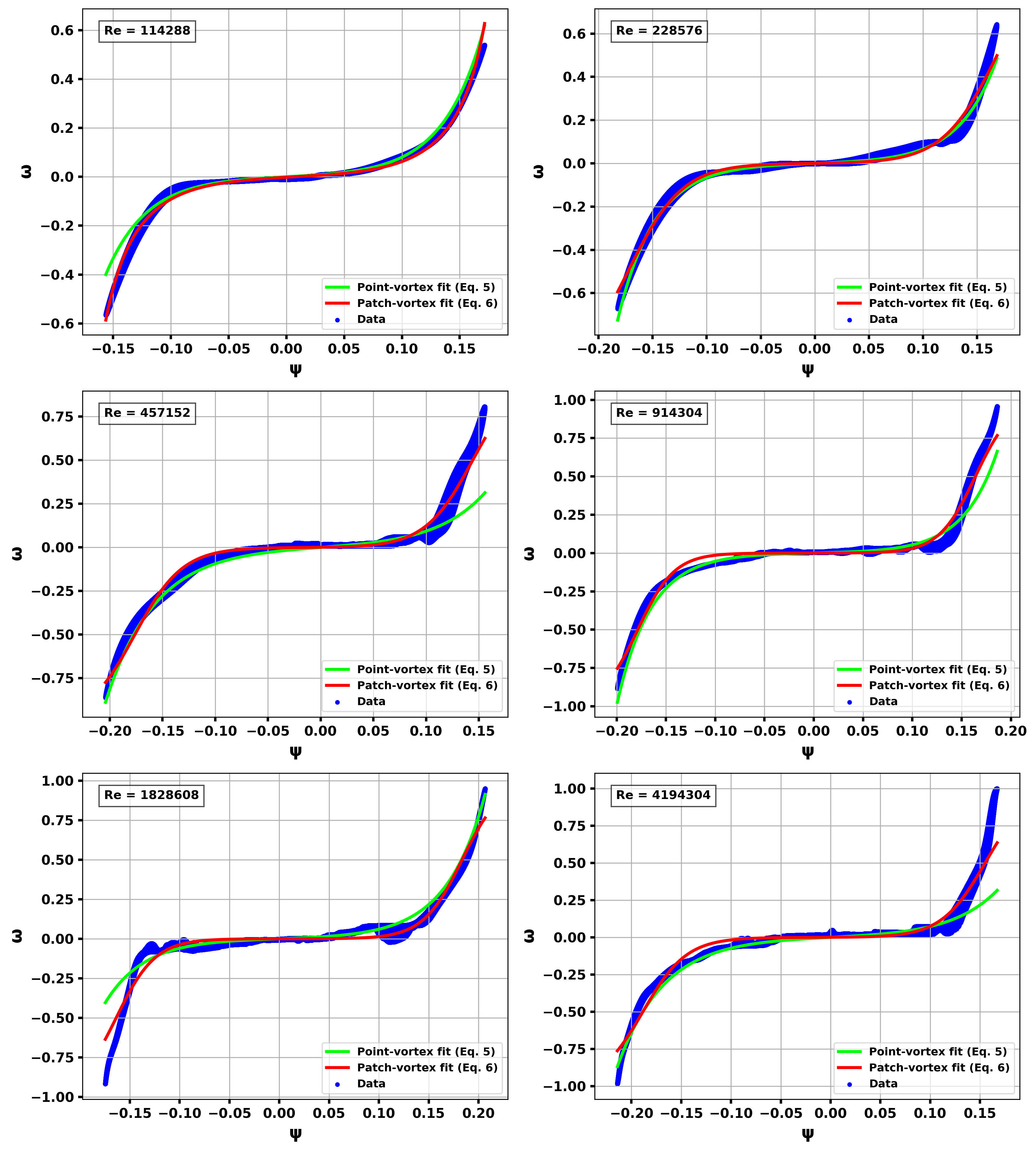}
    \caption{Late-time $\omega$--$\psi$ scatter plots for the highest VPF case ($62.5\%$) across different Reynolds numbers, showing comparison with point-vortex ($\sinh$) \textcolor{red}{[Eq.~(\ref{eq:pv})]} and patch-vortex (KMRS) \textcolor{red}{[Eq.~(\ref{eq:kmrs})]} statistical-equilibrium fits. Results are shown at $T=3000$ for all cases except Re=4194304, which is shown at $T=5000$.}
    \label{fig:ESM_KMRS}
\end{figure*}

The late-time Eulerian statistical equilibria in the $\omega$--$\psi$ plane were shown in MBG2026 to exhibit a transition from point-vortex to finite-size patch-vortex behavior with increasing VPF for Re = 228576. The equilibrium states for the highest VPF case across different Re are presented in Fig.~\ref{fig:ESM_KMRS}. We observe that, at high VPF, the late-time equilibrium state remains well described by the finite-size patch-vortex KMRS theory across the entire Re range. Therefore, the late-time equilibrium statistical state \textcolor{black}{is found} to be governed primarily by the VPF and remains largely unaffected by variations in the Reynolds number. \textcolor{black}{We note that all cases were evolved up to $T=3000$, except for the highest-Reynolds-number case ($\mathrm{Re}=4194304$), which was extended to $T=5000$ to allow remnant small-scale vortices to undergo further long-time interactions with the coherent vortices before ultimately merging or dissipating into the background flow. The corresponding $\omega$--$\psi$ scatter plots at $T=3000$ and $T=5000$ for Re = 4194304 remain qualitatively similar, with the  vertical spike observed at $T=3000$ (not shown here), no longer present at $T=5000$. This spike is associated with remnant small-scale vortices undergoing long-time interactions with the coherent vortices before eventually merging or dissipating into the background flow.
}

\textcolor{black}{Although the kinetic-energy decay rates differ significantly with Reynolds number, reflecting the varying strength of viscous dissipation, the resulting late-time $\omega$--$\psi$ equilibria remain remarkably similar. This indicates that the Reynolds number influences the transient relaxation dynamics but has little effect on the nature of the final Eulerian statistical equilibrium.}

\subsection {Lagrangian Dynamics}

We previously presented the Lagrangian transport for various VPF at a Reynolds number of Re = 228576. Here, we extend the analysis over a two-dimensional (VPF, Re) parameter space.

\subsubsection{Transport across various VPF and RE}

Here, we present the transport results using mean-square displacement (MSD) and diffusion \textcolor{black}{of tracer particles} for each VPF case, with comparisons across multiple Reynolds numbers for every VPF.

\begin{figure*}[t]
    \centering
    \begin{subfigure}[b]{0.49\textwidth}
        \centering
        \includegraphics[width=\textwidth]{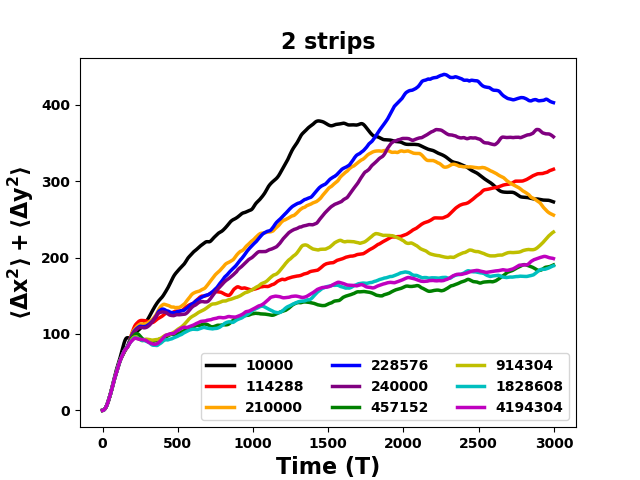} 
        \caption{}
        \label{fig:KE_y}  % <-- after \caption
    \end{subfigure}
    \hfill
    \begin{subfigure}[b]{0.49\textwidth}
        \centering
        \includegraphics[width=\textwidth]{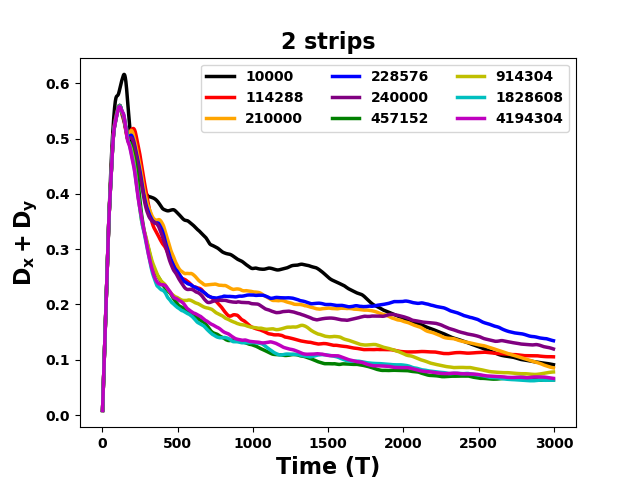}
        \caption{}
        \label{fig:Tr_2}  
    \end{subfigure}
    \caption{ Total transport of tracer particles for the 2-strip initial configuration at varying Reynolds numbers (Re = 10000 – 4194304) shown in linear–linear axes. Different colors represent different Reynolds numbers. (a) Time evolution of the mean square displacement (MSD). (b) Corresponding total diffusion coefficients, $Dx+Dy$.}
    \label{fig:2}
\end{figure*}

The transport behavior for the 2-strip case across different Reynolds numbers is shown in Fig.~\ref{fig:2}. The transport remains predominantly subdiffusive across the investigated Reynolds-number range in both the inverse-cascade turbulent regime and the late-time large-scale coherent-vortex regime. The transport exponent averaged over all Reynolds numbers for the fixed 2-strip VPF configuration, $\langle \alpha_T \rangle_{Re}$, is $0.36$ and $\langle \alpha_C \rangle_{Re}$ is $0.11$, corresponding to subdiffusive transport ($\alpha<1$) in both the inverse-cascade and coherent-vortex regimes, respectively (see Table~\ref{tab:transport_mean}), representing the lowest mean transport levels among all the VPF cases investigated.

\begin{figure*}[t]
    \centering
    \begin{subfigure}[b]{0.49\textwidth}
        \centering
        \includegraphics[width=\textwidth]{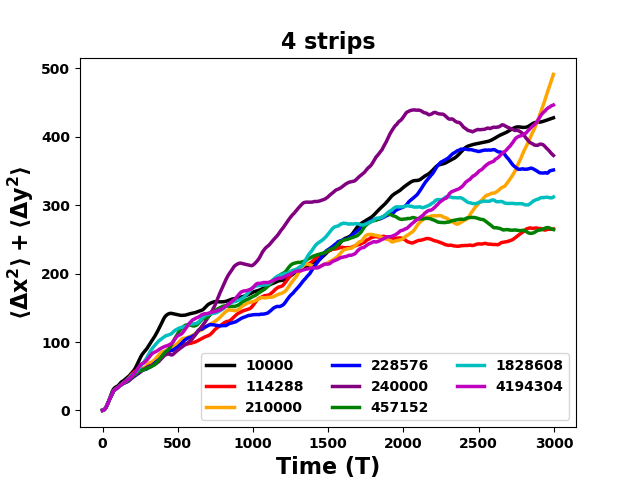} 
        \caption{}
        \label{fig:KE_y}  % <-- after \caption
    \end{subfigure}
    \hfill
    \begin{subfigure}[b]{0.49\textwidth}
        \centering
        \includegraphics[width=\textwidth]{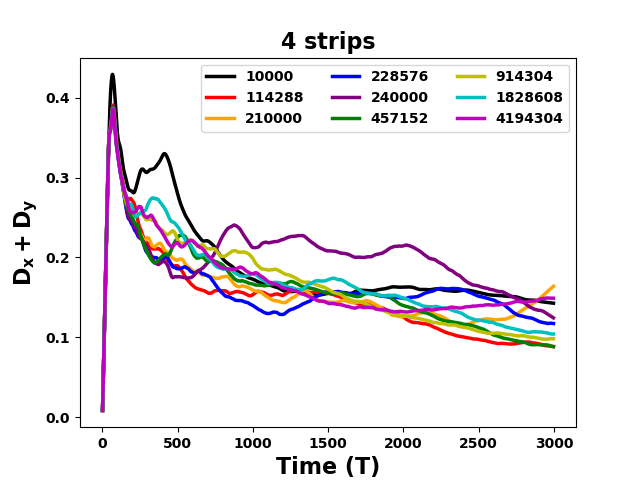}
        \caption{}
        \label{fig:Tr_4}  
    \end{subfigure}
    \caption{ Total transport of tracer particles for the 4-strip initial configuration at varying Reynolds numbers (Re = 10000 – 4194304) shown in linear–linear axes. Different colors represent different Reynolds numbers. (a) Time evolution of the mean square displacement (MSD). (b) Corresponding total diffusion coefficients, $Dx+Dy$.}
    \label{fig:4}
\end{figure*}

The transport behavior for the 4-strip case across different Reynolds numbers is shown in Fig.~\ref{fig:4}. The transport remains predominantly subdiffusive across the investigated Reynolds-number range during the inverse-cascade turbulent regime. In the late-time large-scale coherent-vortex regime, the transport also remains predominantly subdiffusive for most Reynolds numbers, with the exception of Re = 210000, where superdiffusive transport is observed. 
The transport exponent averaged over all Reynolds numbers for the fixed 4-strip VPF configuration, $\langle \alpha_T \rangle_{Re}$, is $0.78$ and $\langle \alpha_C \rangle_{Re}$ is $0.49$, corresponding to subdiffusive transport ($\alpha<1$) in both the inverse-cascade and coherent-vortex regimes, respectively (see Table~\ref{tab:transport_mean}), indicating a slight increase in the mean transport level compared to the 2-strip case.

\begin{figure*}[t]
    \centering
    \begin{subfigure}[b]{0.49\textwidth}
        \centering
        \includegraphics[width=\textwidth]{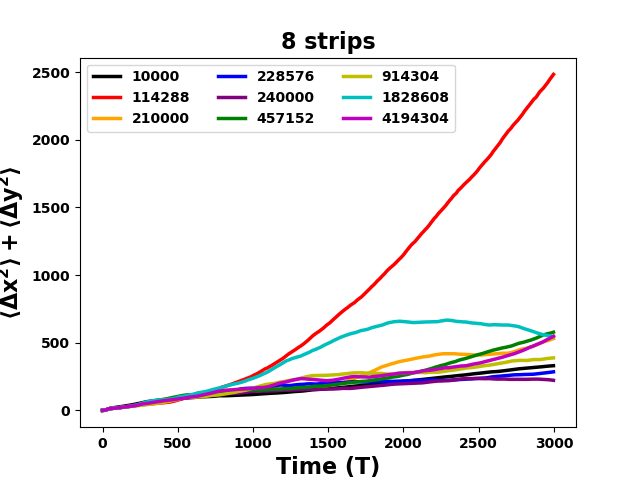} 
        \caption{}
        \label{fig:KE_y}  % <-- after \caption
    \end{subfigure}
    \hfill
    \begin{subfigure}[b]{0.49\textwidth}
        \centering
        \includegraphics[width=\textwidth]{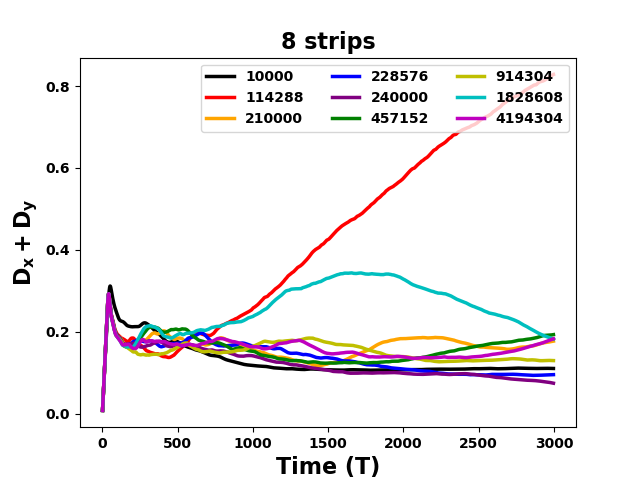}
        \caption{}
        \label{fig:Tr_8}  
    \end{subfigure}
    \caption{ Total transport of tracer particles for the 8-strip initial configuration at varying Reynolds numbers (Re = 10000 – 4194304) shown in linear–linear axes. Different colors represent different Reynolds numbers. (a) Time evolution of the mean square displacement (MSD). (b) Corresponding total diffusion coefficients, $Dx+Dy$.}
    \label{fig:8}
\end{figure*}

The transport behavior for the 8-strip case across different Reynolds numbers is shown in Fig.~\ref{fig:8}. During the inverse-cascade turbulent regime, the transport lies in a transitional regime between subdiffusive and superdiffusive behavior across the investigated Reynolds-number range, with the transport coefficients spanning values close to and slightly greater than unity. The transport exponent averaged over all Reynolds numbers for the fixed 8-strip VPF configuration, $\langle \alpha_T \rangle_{Re}$, is $1.08$ in the inverse-cascade regime, indicating the onset of superdiffusive transport. In the late-time large-scale coherent-vortex regime, the transport oscillates between subdiffusive and superdiffusive behavior for different Reynolds numbers, with the mean transport level across Reynolds numbers remaining close to and slightly greater than unity. The corresponding averaged transport exponent is $\langle \alpha_C \rangle_{Re}=1.05$, representing the highest mean transport level among all the VPF cases investigated. This configuration also exhibits the largest number of superdiffusive realizations across the investigated Reynolds-number range (see Table~\ref{tab:transport_mean}).

\begin{figure*}[t]
    \centering
    \begin{subfigure}[b]{0.49\textwidth}
        \centering
        \includegraphics[width=\textwidth]{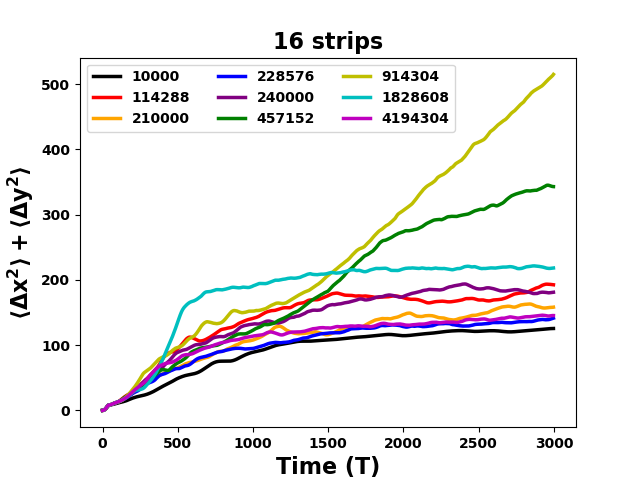} 
        \caption{}
        \label{fig:KE_y}  % <-- after \caption
    \end{subfigure}
    \hfill
    \begin{subfigure}[b]{0.49\textwidth}
        \centering
        \includegraphics[width=\textwidth]{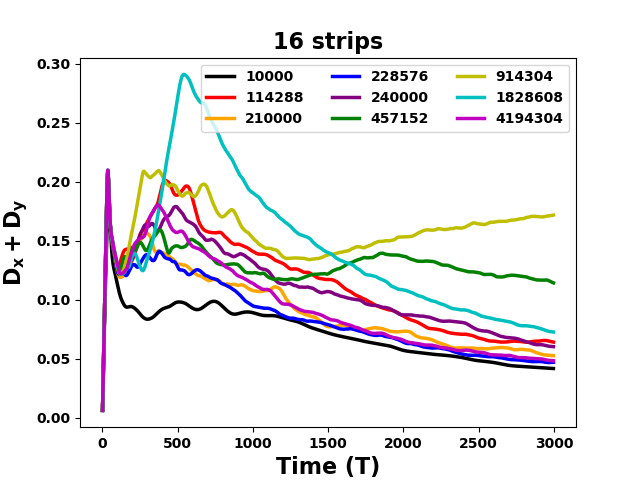}
        \caption{}
        \label{fig:Tr_16}  
    \end{subfigure}
    \caption{ Total transport of tracer particles for the 16-strip initial configuration at varying Reynolds numbers (Re = 10000 – 4194304) shown in linear–linear axes. Different colors represent different Reynolds numbers. (a) Time evolution of the mean square displacement (MSD). (b) Corresponding total diffusion coefficients, $Dx+Dy$.}
    \label{fig:16}
\end{figure*}

The transport behavior for the 16-strip case across different Reynolds numbers is shown in Fig.~\ref{fig:16}. The transport becomes predominantly superdiffusive across the investigated Reynolds-number range during the inverse-cascade turbulent regime. In the late-time large-scale coherent-vortex regime, the transport remains predominantly subdiffusive for most Reynolds numbers, with the exception of Re = 914304, where superdiffusive transport is observed. The transport exponent averaged over all Reynolds numbers for the fixed 16-strip VPF configuration, $\langle \alpha_T \rangle_{Re}$, is $1.25$ and $\langle \alpha_C \rangle_{Re}$ is $0.42$ in the inverse-cascade and coherent-vortex regimes, respectively (see Table~\ref{tab:transport_mean}). This represents a further increase in the mean transport level in the inverse-cascade regime, accompanied by a reduction in the mean transport level in the late-time coherent-vortex regime compared to the 8-strip case.

\begin{figure*}[t]
    \centering
    \begin{subfigure}[b]{0.49\textwidth}
        \centering
        \includegraphics[width=\textwidth]{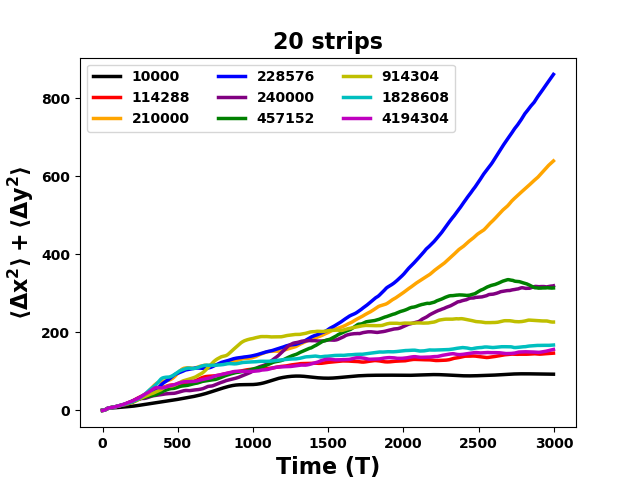} 
        \caption{}
        \label{fig:KE_y}  % <-- after \caption
    \end{subfigure}
    \hfill
    \begin{subfigure}[b]{0.49\textwidth}
        \centering
        \includegraphics[width=\textwidth]{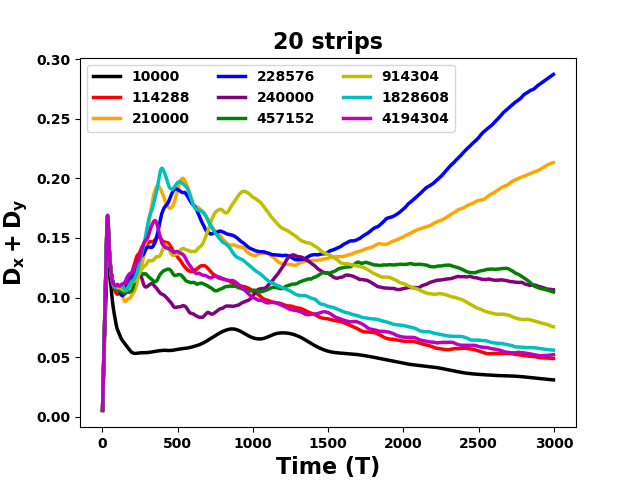}
        \caption{}
        \label{fig:Tr_20}  
    \end{subfigure}
    \caption{Total transport of tracer particles for the 20-strip initial configuration at varying Reynolds numbers (Re = 10000 – 4194304) shown in linear–linear axes. Different colors represent different Reynolds numbers. (a) Time evolution of the mean square displacement (MSD). (b) Corresponding total diffusion coefficients, $Dx+Dy$.}
    \label{fig:20}
\end{figure*}

The transport behavior for the 20-strip case across different Reynolds numbers is shown in Fig.~\ref{fig:20}. The transport remains predominantly superdiffusive across the investigated Reynolds-number range during the inverse-cascade turbulent regime. In the late-time large-scale coherent-vortex regime, the transport remains predominantly subdiffusive for most Reynolds numbers, with the exception of Re = 210000 and Re = 228576, where superdiffusive transport is observed. The transport exponent averaged over all Reynolds numbers for the fixed 20-strip VPF configuration, $\langle \alpha_T \rangle_{Re}$, is $1.35$ and $\langle \alpha_C \rangle_{Re}$ is $0.63$ in the inverse-cascade and coherent-vortex regimes, respectively (see Table~\ref{tab:transport_mean}). This represents a further increase in the mean transport level in the inverse-cascade regime, attaining the highest mean transport among all the VPF cases investigated, while the late-time coherent-vortex regime continues to exhibit an overall subdiffusive mean transport level ($\alpha<1$).

\begin{table}[h!]
\centering
\caption{Reynolds-number-averaged transport exponents during the inverse-cascade turbulent regime, $\langle \alpha_T \rangle_{Re}$, and the late-time coherent-vortex regime, $\langle \alpha_C \rangle_{Re}$, for different vorticity packing fractions (VPF).}
\begin{tabular}{|c|c|c|}
\hline
\textbf{Initial strips} & $\mathbf{\langle \alpha_T \rangle_{Re}}$ & $\mathbf{\langle \alpha_C \rangle_{Re}}$ \\
\hline
2  & 0.36 & 0.11 \\
4  & 0.78 & 0.49 \\
8    & 1.08 & 1.05 \\
16    & 1.25 & 0.42 \\
20  & 1.35 & 0.63 \\
\hline
\end{tabular}
\label{tab:transport_mean}
\end{table}

To facilitate comparison across the (VPF, Re) parameter space, the transport exponents $\alpha_T$  and $\alpha_C$  are presented as line plots in Figs.~\ref{fig:tr_inversecasc} and  \ref{fig:tr_coherent}, respectively. The corresponding numerical values of these transport exponents are summarized in the heat-map representation shown in Appendix Fig.~\ref{fig:Heatmap}.

\subsubsection{ Transport in the inverse-cascade regime}

\begin{figure*}[t]
    \centering
    \includegraphics[width=\textwidth]{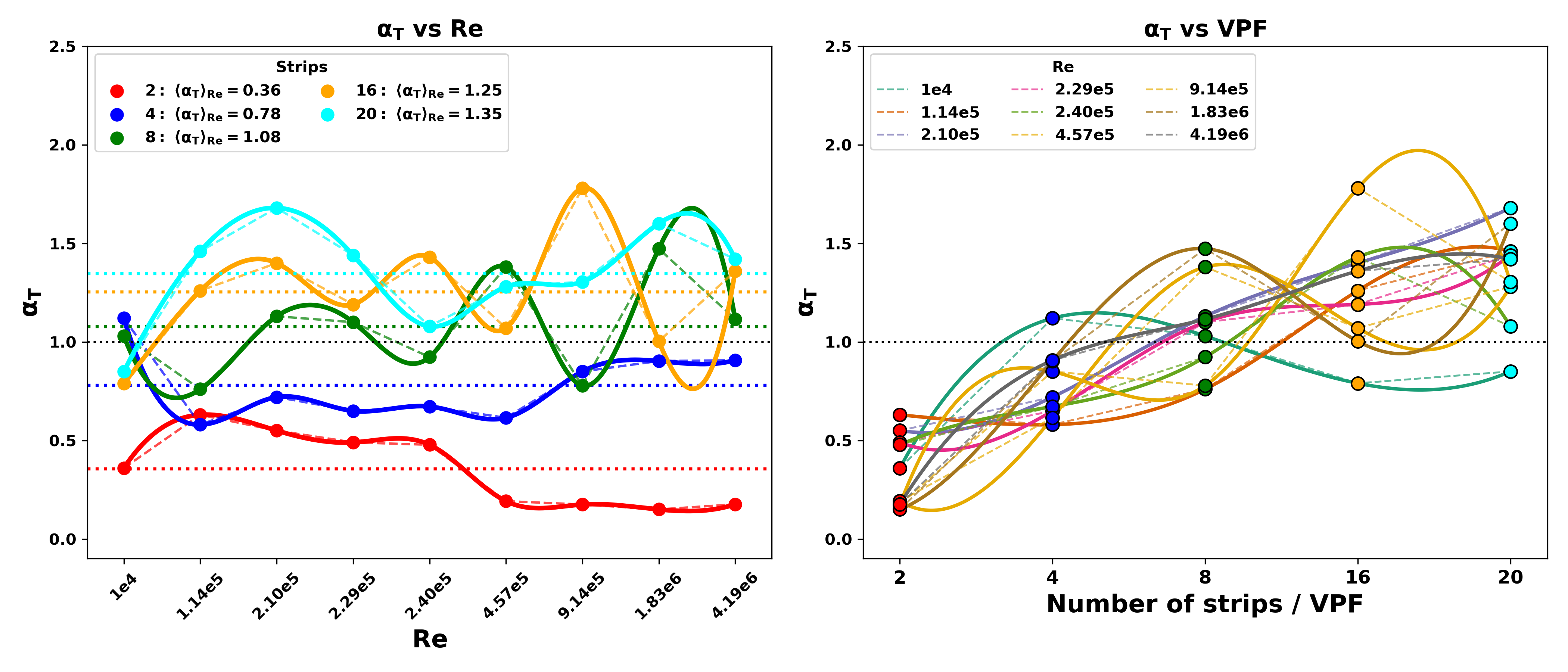}
    \caption{Transport exponent $\alpha_T$ in the inverse-cascade regime across the two-dimensional (VPF, Re) parameter space. (a) Variation of $\alpha_T$ with Re for different vortex packing fractions (VPF). Horizontal dotted lines denote the mean transport level  $\langle \alpha_T \rangle_{\mathrm{Re}}$ for each VPF averaged across all Reynolds numbers. (b) Variation of $\alpha_T$ with VPF for different Reynolds numbers.}
    \label{fig:tr_inversecasc}
\end{figure*}

 During the inverse-cascade regime, the transport exponent exhibits an overall increasing trend with increasing VPF, while Re introduces secondary oscillatory modulation around the mean transport behavior. This trend becomes more evident from the horizontal dotted lines in Fig.~\ref{fig:tr_inversecasc}(a), which represent the mean transport level obtained by averaging each transport curve across all Reynolds numbers. The mean transport level increases systematically and nearly monotonically with increasing VPF. As summarized in Table~\ref{tab:transport_mean}, the Reynolds-number-averaged transport exponent increases from $\langle \alpha_T \rangle_{\mathrm{Re}} = 0.36$ and $0.78$ for the 2- and 4-strip cases, respectively, to $1.08$, $1.25$, and $1.35$ for the 8-, 16-, and 20-strip cases, respectively.
In particular, the low-VPF cases (2 and 4 strips) remain predominantly subdiffusive, whereas the intermediate- and high-VPF cases (8, 16, and 20 strips) progressively transition toward superdiffusive transport. Increasing VPF is associated with increasingly vigorous turbulent dynamics, resulting in a transition from predominantly subdiffusive to superdiffusive behavior. At the same time, for a fixed VPF, the transport exponent exhibits  oscillatory and weakly non-monotonic variation with Reynolds number. Nevertheless, despite these variations, the Reynolds-averaged transport level remains strongly ordered by VPF, indicating that VPF remains the primary parameter governing the overall transport hierarchy in the inverse-cascade regime, while Reynolds number acts primarily as a secondary modulation. This conclusion is further supported by Fig.~\ref{fig:tr_inversecasc}(b), which shows that, for a fixed Reynolds number, the transport generally increases from low- to high-VPF states despite the presence of local non-monotonic variations.

\subsubsection{Transport in the coherent-vortex regime}

\begin{figure*}[htbp]
    \centering
    \includegraphics[width=\textwidth]{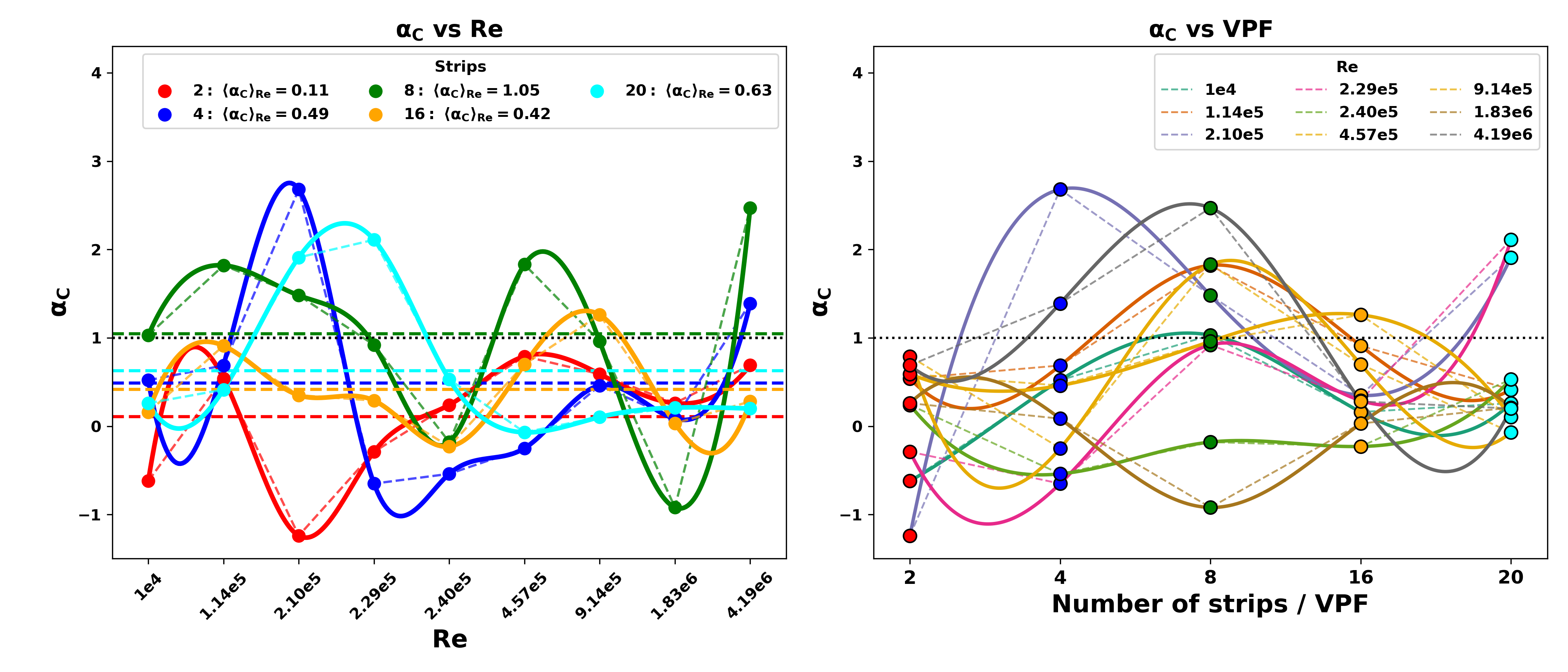}
    \caption{Transport exponent $\alpha_C$ in the coherent vortex regime across the two-dimensional (VPF, Re) parameter space. (a) Variation of $\alpha_C$ with Re for different vortex packing fractions (VPF). Horizontal dotted lines denote the mean transport level $\langle \alpha_C \rangle_{\mathrm{Re}}$ for each VPF averaged across all Reynolds numbers. (b) Variation of $\alpha_C$ with VPF for different Reynolds numbers.}
    \label{fig:tr_coherent}
\end{figure*}

In contrast to the inverse-cascade regime, the transport behavior in the coherent dipole stage becomes strongly oscillatory and highly non-monotonic across both Re and VPF. The transport exponent spans a broad range from subdiffusive to strongly superdiffusive regimes, indicating strong sensitivity of long-time transport to the detailed coherent-vortex organization and dipole interaction dynamics. The observed subdiffusive behavior is associated with coherent dipolar vortices undergoing confined orbital motion, resulting in enhanced particle trapping both within the vortex cores and around the orbiting dipolar structures. In contrast, the superdiffusive behavior is associated with the formation of \textcolor{black}{long-lived dipolar states undergoing persistent translational motion}.

For a fixed VPF, $\alpha$  exhibits strong oscillatory variation with Re while for a fixed Reynolds number the dependence on VPF also remains oscillatory, highly irregular, and without any simple monotonic ordering.  This behavior is evident in Fig.~\ref{fig:tr_coherent}(a), where the Reynolds-number-averaged transport levels, indicated by the horizontal dotted lines, no longer exhibit the monotonic increase with VPF observed in Fig.~\ref{fig:tr_inversecasc}(a), but instead display a broad maximum near the intermediate 8-strip configuration. Furthermore, Fig.~\ref{fig:tr_coherent}(b) shows that individual Reynolds-number realizations exhibit strongly irregular variation of transport with VPF, in contrast to the broadly increasing VPF-dependent trend observed during the inverse-cascade regime. Nevertheless, despite these strong local oscillations, the Reynolds-number-averaged transport level remains largest for the intermediate 8-strip configuration.  As summarized in Table~\ref{tab:transport_mean}, the Reynolds-number-averaged transport exponent increases from $\langle \alpha_C \rangle_{\mathrm{Re}}=0.11$ and $0.49$ for the 2- and 4-strip cases, respectively, to a maximum value of $1.05$ for the 8-strip case, before decreasing to $0.42$ and $0.63$ for the 16- and 20-strip cases, respectively. Transport remains subdiffusive for all Re in the 2-strip case and for most Re in the 4-, 16-, and 20-strip cases, with only occasional superdiffusive states appearing at isolated Re. The intermediate 8-strip configuration is particularly notable, exhibiting the largest number of superdiffusive instances and the highest mean transport level. 

A notable exception occurs for Re=228576, where our earlier work MBG2026 demonstrated an approximately monotonic increase of the transport exponent with VPF in the late-time coherent-vortex equilibrium regime, thereby establishing a direct Eulerian–Lagrangian correspondence between the transition of the Eulerian statistical equilibrium states and the associated tracer transport properties. Although weak oscillatory modulation remains present across the individual VPF states, the overall transport level increases from predominantly subdiffusive low-VPF states toward strongly superdiffusive behavior at higher VPF. Such a systematic VPF-dependent enhancement is not observed consistently across the other Reynolds numbers investigated here.

\textcolor{black}{It should be noted that the present transport statistics are obtained over a finite simulation duration of $T=3000$. Since the late-time coherent-vortex regime is characterized by long-lived vortex interactions and reorganizations, the detailed transport coefficients may continue to evolve over longer integration times, \textcolor{black}{albeit relatively slowly.} Consequently, some of the quantitative values of $\alpha_C$, as well as the occurrence of individual subdiffusive or superdiffusive realizations, may be modified at sufficiently long times. Nevertheless, the present results demonstrate that, within the simulation interval investigated here, the coherent-vortex transport exhibits substantially stronger Reynolds-number-dependent variability than that observed during the inverse-cascade regime.} 

The broader (VPF, Re) scan suggests that the previously observed direct Eulerian–Lagrangian correspondence is not universally preserved across the entire parameter space. While the late-time Eulerian statistical equilibria continue to exhibit a robust transition from point-vortex to KMRS-like states with increasing VPF, the associated Lagrangian transport properties display strong Reynolds-number-dependent variability, spanning subdiffusive to superdiffusive regimes. Unlike the inverse-cascade regime, where Reynolds-number variations primarily modulate an underlying VPF-ordered transport hierarchy, the coherent-vortex regime exhibits sufficiently strong Reynolds-number-dependent oscillations to substantially weaken this hierarchy. This indicates that the long-time transport dynamics are influenced not only by the underlying Eulerian equilibrium class, but also by the detailed coherent-vortex dynamics and dipole organization.

\section{Conclusions}

In this manuscript, we \textcolor{black}{present} results from numerical simulations of decaying two-dimensional incompressible Navier–Stokes turbulence driven by Kelvin–Helmholtz instability, together with coupled passive tracer-particle dynamics, investigating the effects of varying initial vorticity packing fraction (VPF) and Reynolds number (Re), on the associated Eulerian–Lagrangian connections across the inverse-cascade and coherent-vortex regimes. \textcolor{black}{Our main findings are summarized below:} \\

1) The late-time quasi-steady Eulerian statistical equilibria exhibit a transition from point-vortex to finite-size patch-vortex (KMRS) behavior with increasing VPF, \textcolor{black}{as demonstrated  in MBG2026.} For a fixed VPF, variations in Re do not alter the late-time equilibrium states. The highest-VPF case presented here remains consistently well described by the KMRS vorticity–streamfunction relation across the entire Re range investigated. Although the kinetic-energy decay rates vary significantly with Re due to differences in viscous dissipation, these differences primarily affect the transient relaxation dynamics and have little influence on the resulting late-time Eulerian statistical equilibrium.

2) In the inverse-cascade turbulent regime, the transport exponent increases systematically and nearly monotonically with increasing VPF. For a fixed VPF, variations in Reynolds number introduce  oscillatory modulation of the transport coefficients around the mean transport level. Nevertheless, the Reynolds-number-averaged transport remains strongly ordered by VPF, indicating that transport in this regime is governed primarily by the VPF, while Reynolds number acts mainly as a secondary modulation.

3) In the late-time coherent-vortex regime, the transport exponent becomes strongly dependent on both VPF and Reynolds number, exhibiting highly oscillatory and non-monotonic behavior across the two-dimensional (VPF, Re) parameter space. Unlike the inverse-cascade regime, the Reynolds-number-dependent oscillations substantially weaken the VPF-ordered transport hierarchy, resulting in transport behavior that ranges from strongly subdiffusive to strongly superdiffusive states.

4) In the nonequilibrium inverse-cascade regime, the broad Eulerian–Lagrangian relationship remains preserved across the (VPF, Re) parameter space. However, the direct late-time  Eulerian–Lagrangian correspondence previously observed at Re=228576, connecting the transition from point-vortex to patch-vortex statistical equilibria with the transition from subdiffusive to superdiffusive transport with increasing VPF, is not universally preserved across the broader Reynolds-number range investigated here. \textcolor{black}{This suggests that, although the late-time Eulerian equilibrium state remains primarily controlled by the VPF, the associated Lagrangian transport properties depend additionally on Reynolds-number-sensitive features of the coherent-vortex dynamics that are not fully captured by the equilibrium $\omega$--$\psi$ relation alone. Unlike the inverse-cascade regime, where Reynolds-number variations primarily modulate an underlying VPF-ordered transport hierarchy, the coherent-vortex regime exhibits sufficiently strong Reynolds-number-dependent oscillations to substantially weaken this hierarchy. This indicates that the long-time transport dynamics are influenced not only by the underlying Eulerian equilibrium class, but also by the detailed coherent-vortex dynamics and dipole organization.}

\begin{acknowledgments}
The simulations and visualizations presented here are
performed on GPU nodes and visualization nodes of the ANTYA cluster at the Institute for Plasma Research (IPR), India. The authors are grateful to the HPC support team of IPR for extending their help related to the ANTYA cluster.
\end{acknowledgments}

\nocite{*}
\bibliography{References}% Produces the bibliography via BibTeX.

\onecolumngrid
%\clearpage
\appendix
\section{\textbf{}}\label{Appen}

\begin{figure*}[htbp]
    \centering
    \includegraphics[width=0.95\textwidth]{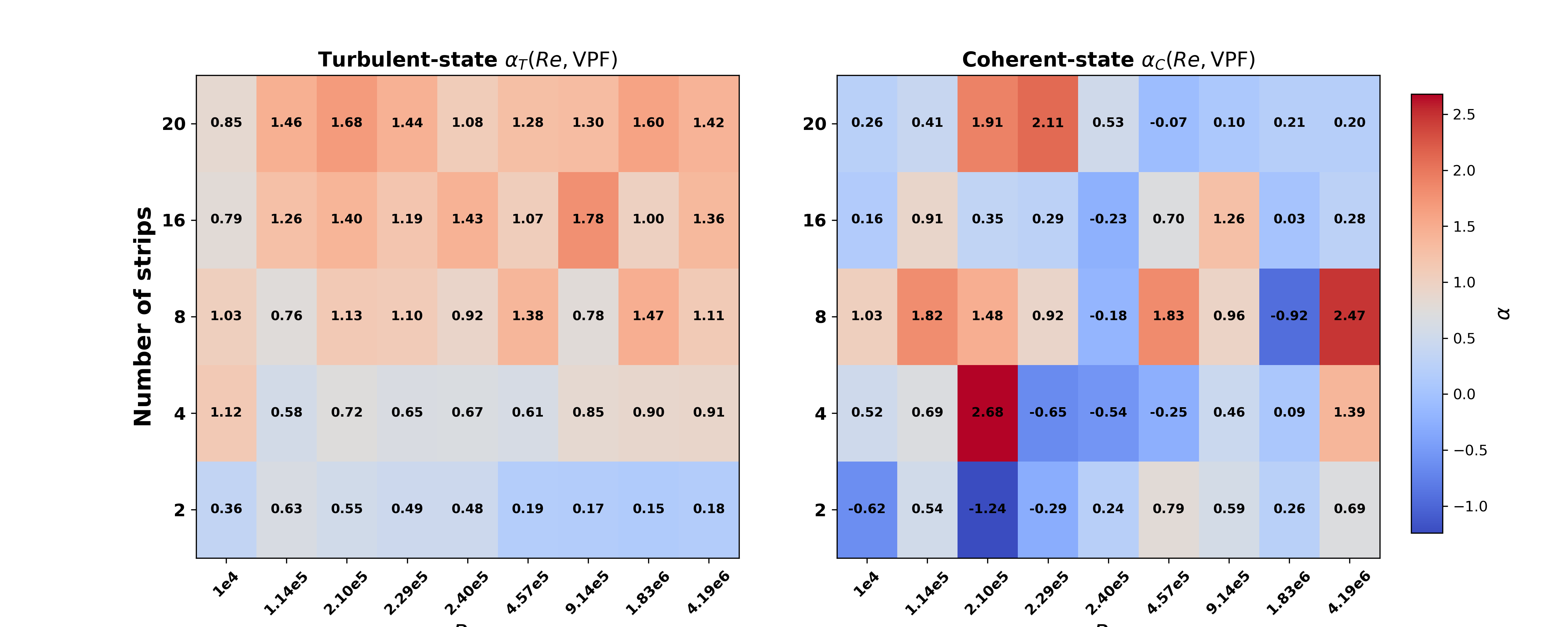}
    \caption{ Heat-map representation of the transport exponents $\alpha_T$  (inverse-cascade regime) and $\alpha_C$  (coherent-vortex regime) across the two-dimensional (VPF, Re) parameter space.  
    }
    \label{fig:Heatmap}
\end{figure*}

The transport coefficients obtained from Lagrangian tracer transport for various combinations of VPF and Reynolds number across both the inverse-cascade turbulent regime and the late-time quasi-equilibrium coherent-vortex regime are presented as heat maps in Fig.~\ref{fig:Heatmap}.

\end{document}